\documentstyle[aps,twocolumn,prl]{revtex}

\begin{document}
 
\newcommand{\lsim}{\raisebox{1.5pt}{\small $<$}\hspace*{-6.7pt}\raisebox{-
3pt}{\small $\sim$} }
\newcommand{\gsim}{\raisebox{1.5pt}{\small $>$}\hspace*{-6.7pt}\raisebox{-
3pt}{\small $\sim$} }

\draft

\title{Specific heat study of single crystalline \mbox{Pr$_{0.63}$ Ca$_{0.37}$ 
MnO$_{3}$} in presence of a \mbox{magnetic field}}
\author{A.K.Raychaudhuri $^\dagger$ and Ayan Guha}
\address{Department of Physics, Indian Institute of Science, 
Bangalore 560 012, India}
\author{I.Das and R. Rawat}
\address{Inter University Consortium for DAE facilities, University Campus,
Khandwa Road, \mbox{Indore- 452017, India}}
\author{C.N.R. Rao}
\address{CSIR Center of Excellence in Chemistry, Jawaharlal Nehru Center for
Advanced Scientific Research, Jakkur P.O., \mbox{Bangalore 560 064,} India}

\date{\today}

\twocolumn[\hsize\textwidth\columnwidth\hsize\csname @twocolumnfalse\endcsname

\maketitle
\begin{abstract}
We present the results of a study of specific heat on a single crystal of
Pr$_{0.63}$Ca$_{0.37}$MnO$_3$ performed over a temperature range 3K-300K
in presence of 0 and 8T magnetic fields. An estimate of the entropy and 
latent heat in a magnetic field at the first order charge ordering (CO) 
transition is presented. The total entropy change at the CO transition 
which is $\approx$ 1.8 J/mol K at 0T, decreases to $\sim$ 1.5 J/mol K in 
presence of 8T magnetic field. Our measurements enable us to estimate the
latent heat $L_{CO}$ $\approx$ 235 J/mol involved in the CO transition.
Since the entropy of the ferromagnetic metallic (FMM) state is comparable
to that of the charge-ordered insulating (COI) state, a subtle change in
entropy stabilises either of these two states. Our low temperature 
specific heat measurements reveal that the linear term is absent in 0T 
and surprisingly not seen even in the metallic FMM state.  

\end{abstract}

\pacs{75.30.Kz  65.40.+g  75.30.Vn}
]

The discovery of a number of fascinating properties like colossal magneto-
resistance, charge / orbital ordering and electronic phase separation in 
manganites with generalized formula $Re_{1-x}Ae_{x}MnO_{3}$ ($Re$ being a 
trivalent rare-earth ion, $Ae$ a divalent alkaline earth element) has 
resulted in a spurt of research activities ~\cite{Kuwahara,Raorev}.
For certain values of x, close to 0.5, these manganites undergo a
first order transition at certain temperature $T_{CO}$ to a Charge Ordered 
Insulating (COI) state where the Mn$^{3+}$ and Mn$^{4+}$ species arrange 
themselves in a commensurate order in the lattice.
The charge ordering transition in these oxides is accompanied by a large 
change in volume and hysteresis in resistivity and is believed to be a first 
order transition. A fascinating aspect of the COI state (which is also 
accompanied by orbital ordering) is that it is unstable under an applied 
magnetic field and there is an insulator- metal transition (melting) of 
the COI state to a ferromagnetic metal state (FMM) below a temperature 
$T_{MH}$.

In this paper we have investigated the specific heat and related 
thermodynamic quantities in a single crystalline CO system over a wide 
temperature range ($3K < T < 300K$) and in a magnetic field upto 8T which 
can melt the CO order. A study of specific heat over an extensive 
temperature range is crucial in understanding the nature of the CO 
transition and in addition it can provide values of various fundamental 
parameters of manganites like the density of states at the Fermi level 
$N(E_{F})$, the Debye temperature $\theta_D$, ferromagnetic / 
antiferromagnetic spin wave stiffness constant etc.

Measurements of specific heat in manganites particularly with CMR composition 
has been reported before, particularly in the low temperature region. In the 
CO systems there are reports of specific heat measurements both in single 
and polycrystalline samples mainly at low temperatures. Specific heat 
measurements in CO systems near $T_{CO}$ have been done in polycrystalline 
samples and no magnetic field data have been reported. There exists no data 
on thermodynamics associated with the transition at $T_{MH}$.

The charge ordering transition in these oxides is believed to be a first order 
transition, as mentioned before. Measurements across the CO transition 
should therefore show a discontinuous jump in entropy as the compound absorbs 
latent heat from the bath to transform to a new phase. However, till date there 
exists no proof from calorimetry that the transitions are indeed first order and 
whether a latent heat is released at the transition. There has been no calorimetric investigation of the melting of COI state to a FMM state, the transition which is 
also believed to be a first order transition.

In this paper we investigate this fundamental issues related to changes in 
entropy and latent heat across the charge-ordering transition near $T_{CO}$ 
and $T_{MH}$ in a single-crystal of Pr$_{0.63}$Ca$_{0.37}$MnO$_3$. Pr$_{1-x}$
Ca$_{x}$MnO$_3$ happens to be a prototypical and perhaps the most studied 
charge-ordered system. Due to its low tolerance factor it remains insulating 
for all values of x. For $x = 0.37$ composition, charge and orbital ordering 
occurs at $T_{CO}$ ($\approx$ 235K), while a long range AFM order sets in 
only below $T_N$ ($\approx$ 170K). A small ferromagnetic component appears 
with the canting of AFM spins at a lower temperature, $T_{CA}$ ($\approx$ 30K).

Though there have been a number of studies on low temperature specific heat of 
this particular system ($T < 20K$), a clear picture is yet to be emerge on the 
presence (or absence) of a linear term in specific heat. Recently there have 
been reports of large linear contribution to specific heat and appearance of 
excess specific heat associated with charge ordering ~\cite{SMOLN} in this system. 
Since such a linear term is often associated with a electronic contribution to 
the specific heat, appearance of this term in an insulating sample is intriguing. 
In this paper, we have also investigated the low temperature region and reached 
certain definite conclusions about the linear term.

In our experiments we have specifically asked the following questions:
\begin{enumerate}
\item Is it possible to identify the relevant transitions in zero and finite magnetic 
fields through calorimetric measurements ?
\item What latent heat is released and the entropy change across the first order 
transitions, one at $T_{CO}$ in zero magnetic field and the other at $T_{MH}$ in 
8T magnetic field ?
\item When the FMM state is obtained from the COI state in presence of a magnetic 
field, do we obtain a linear term (arising from electronic contribution) in the 
specific heat ? 
\item Do we see a linear term in specific heat in COI state as reported by some 
investigators ?
\end{enumerate}

The remainder of the paper is divided into two principal sections. In the first 
section after the experimental section, we present and discuss the experimental 
data in the region $T \geq  50K$ which essentially encompasses the region where 
most of the transition occurs and this also happens to be the region where no 
experimental data have been reported in the past either in presence of magnetic 
field or in a single crystal. The second major section refers to the data at low 
temperatures where issues like linear term in the heat capacity, the Debye term, 
spin wave contributions etc. are looked into. In this region there are past 
studies , as mentioned earlier and we compare our results on these material with 
results from other CO systems.

\section{Experimental}
The crystal used in our experiment has been grown by float zone technique and has 
been used in a number of previous experiments by our group ~\cite{Ayan1,Ayan2}. 
The resistivity ($\rho$) vs. T curve in zero field and in a field of H = 8T are 
shown in fig.~1. The COI state can be melted to a FMM state by application of 
a magnetic field of 8T at $T_{MH} \approx$ 90K. In presence of magnetic field 
the region $T_{MH} < T < T_{CO}$ 
is termed as ``mixed charge order" (MCO) region, where the COI phase coexist with 
the FMM phase. We have used a semi-adiabatic heat pulse technique to measure the 
specific heat in a wide temperature range of 2K - 300K and in  0T and 8T magnetic 
fields.

\section{Specific heat at high ($T\geq 50K$)}
In figure~2 we show the specific heat ($C_p$) data over an extended temperature 
range 3K - 300K for both 0 and 8T. The most prominent feature is the sharp peak at 
$T_{CO}$ both for H = 0 and 8T. The peak is much larger and sharper compared to that 
seen in polycrystalline materials ~\cite{LEES}(which is also plotted in the same graph). 
The peak at $T_{CO}$ retains its narrowness and sharpness at H = 8T while it shifts 
to lower T with $dT_{CO}/dT$ $\approx$ -1K/T.  In figure~3 we show the region close 
to $T_N$. The graph shows the actual observed step like feature in $C_p$ (at H = 0T)
at $T_N$. In the same graph we show the $C_{exc}$ after subtraction of the lattice 
contribution (the procedure for subtraction of the lattice contribution is given in 
the next sub-section). A peak in $C_{exc}\approx 6J/mole K$ at $T_N$ is visible. 
This feature is suppressed by application of a magnetic field of 8T. The entropy 
change, $\Delta S_{exc}(T_N)$, associated with this transition has been estimated 
to be $\approx 0.5-0.8 J/mole.K$.   
Lee {\it et. al.} ~\cite{LEES} reported for $Pr_{0.6}Ca_{0.4}MnO_3$(polycrystal) 
$\Delta S_{exc} (T_N)$ $\approx$ 0.6J/mole K at $T_N$ =160K. Ramirez {\it et.al.}
~\cite{Ramirez} for $La_{0.35}Ca_{0.65}MnO_3$(polycrystal) obtained $\Delta 
S_{exc}(T_N)$ $\approx 1J/mole K$. It can be seen that all these are much less 
than that what one would expect from a complete spin ordering. The calorimetry 
data therefore points towards an incomplete spin ordering at $T_N$.

Another very interesting feature in our calorimetry data is a clear signature 
of the magnetic field induced transition at $T_{MH}$ $\approx$78K 
(see fig.~2 and 9). This small yet distinct feature, discussed in detail later 
on, is the first signature of field-induced melting in a calorimetry experiment. 

\subsection{Estimation of the Lattice contribution}
In this subsection we describe the procedure to estimate the lattice contribution  
to specific heat, the background on which the specific heat contribution by the 
other degrees of freedom add up. Proper estimation of the lattice contribution 
will thus allow us to get the contribution of CO and magnetic ordering to $C_{p}$.
We define $C_{exc}$ as :
\begin{equation}
C_{exc} = C_{p} - C_{lattice}
\end{equation}

According to Dulong and Petit's law, the limiting heat capacity at high 
temperature for a compound with {\it r} atoms per molecule is expected to be 
3{\it r}R, where R is the gas constant ~\cite{Esr}. In our case ($r$ = 5) this 
limiting value of the lattice (vibrational) heat capacity turns out to be 
$\approx$ 125 J / mole K. From our data we find that at $T = 300K$, observed 
$C_p$ is $\approx$ 90$\%$ of the Dulong Petit value. For most published data
on manganites, $C_p$ at room temperature reaches this value $\sim$ 100 -120 
J / mole K. This signifies that bulk of the contribution indeed comes from 
the vibrational heat capacity in this temperature range.  This contribution 
must be subtracted from the observed data to obtain the excess specific heat.

We have obtained the lattice contribution $C_{lattice}$ by fitting $C_p$ in a 
region $40K \le T \le 150K$ using three standard models as described below. 
The calculated $C_{lattice}$ is then extrapolated to $T > 150K$. Since, we are 
interested in $C_p$ in the range 80K - 300K, the lower temperature for lattice 
contribution has been limited only to 40K. The upper limit of 150K was considered 
since it was lower than the AFM and CO transition temperatures. We have not 
estimated the background by including the data about $T_{CO}$. We 
are of the opinion that there is structural transition/ modification 
associated with the CO transition . As a result it is not advisable to include 
the data above $T_{CO}$ in the estimation of lattice heat capacity below $T_{CO}$.

The following models were used for estimating the background lattice 
contribution :\\
The Einstein model ~\cite{Esr}:\\
$C_{Einstein} = 3rR \sum_i a_i [{x_i}^2 e^{x_i} / (e^{x_i} - 1)^2]$ ; where
$x_i = h\nu_E / K_BT_i$. In this model, all the {\it 3rN} independent oscillators 
populate three optical modes in ratios $a_1:a_2:a_3$ having Einstein frequencies 
$h \nu_{E_i}$. The best fit to the data is given the three Einstein modes with 
$h \nu / K_B$ = 145 K, 410 K and 625 K. It is interesting to note that a recent 
Raman measurement on x = 0.37 composition observed optical modes at $\approx$ 
360 K, 417 K, and 648 K. The 417K and 648K modes nicely matches with those 
seen in our specific heat measurements.~\cite{Rajeev}  \\
The Debye model ~\cite{Esr}:\\
$C_{Debye} = 9rR / x_D^3 \int_0^{x_D} x_D^4 e^{x_D} / (e^{x_D} - 1)^2 dx_D$ ; 
where $x_D = h\nu_D / K_BT$. The specific heat is due to collective low-frequency
oscillations of phonons with a cut off frequency given by $h\nu_D / K_B$. For our 
sample, the best fit was obtained with $\theta_D = 470K$. This is typically 
the value of $\theta_D$ seen in most oxides.\\
The Thirring model ~\cite{Thirring}:\\
$C_{Thirring} = 3rR \sum_{n = 0}^{\infty} B_n u^{-n}$ ; where $u = [(T/T_b)^2 
+ 1]$ The harmonic portion of the lattice specific heat can be expressed in a 
series with the above form where $T_b \approx \theta_D / 2\pi $ ~\cite{Gordon}. 
The above expansion permits the harmonic portion of the lattice specific heat 
to be fitted reasonably well down to temperatures $\sim 50 K$ even when the 
Debye temperature is $\sim 500 K$. In our case, $T_b = 65 K$ and we used $n$ 
upto 50. The $T_b$ gives an estimated $\theta_D \approx 410K$, which is close 
to that obtained from fitting the data with Debye model.\\

In figure~4 we have shown the deviation of the observed data from the 
fit (i.e. $\Delta C / C = (C_{obs} - C_{calc})/C_{obs}$) in order to 
ascertain the extent of uncertainties involved in the background 
subtraction. We find that for $T < 100K$ the Debye model shows large 
systematic deviation. The Thirring model shows a systematic deviation 
above 150K (not shown in graph). We find that over the whole range the 
least uncertainty is shown by the Einstein model (maximum deviation from 
fit $\pm 5 \%$) and this is a random deviation.

We have used the same $C_{Lattice}$ for both H = 0 and H = 8T. We found 
that if we fit the $C_p$ at H = 8T from 50K to 140K to the above models as 
we have done for the $C_p$ at H = 0T case, we end up in getting essentially 
the same $C_{Lattice}$ using the Einstein model. There is a small 
($ < 10\%$) systematic deviation (for the Einstein model) which is not
numerically significant to affect our results. This is shown in the inset 
of figure~4. For compounds containing $Pr^{3+}$ $C_P$ may contain a contribution 
coming from the crystal field. We have estimated this crystal field contribution 
or $C_{xtal}$ the crystal field data available for iso-structural compounds 
$PrNiO_3$, $PrGaO_3$. We find that in the range of interest 
($100K < T < 300K$) this crystal field contribution $< 7 \%$ at 100 K and 
is $< 2 \%$ at 300K ~\cite{Rosenkranz}.

\subsection{$C_{exc}$}
In figure~5 we have plotted the value of $C_{exc}$ for both H = 0 and 
H = 8T for $T > 120K$. The data show the $C_{exc}$ using all the three 
models. The $C_{exc}$ data in the region close to $T_{CO}$ is truncated 
in figure 5, because in this region the $C_{exc}$ is very large. 
In this temperature range the differences in the $C_{exc}$ obtained 
after the subtraction of the lattice contribution is 
well within 1 - 2 J/mole K for the Debye and the Einstein model. The 
Thirring model shows systematic deviation for $T > 150K$. Considering all 
the models we find that near $T_N$ the uncertainty is the largest since 
$C_{exc}$ is low and this can be as large as $\pm 25\%$. {\it However 
close to the  CO transition  when $C_{exc}$ shots upto $\approx$ 
220J/mole.K. The uncertainty in the background estimation falls 
below $\pm 2\%$.}
The $C_{exc}$ data also show clearly that barring the small region around    
$T_N$ (where $C_{exc}$$\approx 10\%$ of $C_p$) and near $T_CO$ (where 
 $C_{exc}$$> 50\%$ of $C_p$) the contribution by other degrees of 
freedom compared to the lattice contribution is negligible in the 
temperature range $T > 50K$.

\subsection{Entropy change near $T_{CO}$ in H = 0 and H = 8T}

A phase transition is signaled by a singularity in a thermodynamic potential 
such as free energy. If there is a finite discontinuity in one or more of 
the first derivatives of the free energy, the transition is first 
order. At a first order transition one expects a discontinuous jump in the 
entropy. At the charge ordering transition, (which is conjectured as first 
order transition based on observed hysteresis in transport data on 
field and temperature cycling), our system is expected to absorb the latent 
heat of transformation to transform from charge-ordered phase to a 
charge-disordered phase as it is heated through $T_{CO}$. The temperature should 
remain constant till this process is complete and the entropy change is given as : 
\begin{equation}
\Delta S_{21} = S_2 - S_1 = L_{21} / T 
\end{equation}
where $S_1$,$S_2$ are the entropies of phases 1 and 2, $L_{21}$ is the latent 
heat associated with the phase transformation. At a first order transition, 
since the entropy changes discontinuously, specific heat is actually 
undefined. Because at the transition point, the temperature does not change 
when heat is applied one would expect a $\delta$ function like behaviour. In 
reality, however, one rarely sees such a $\delta$ function like behaviour of 
$C_p$ because the transition can get broadened either by the process of 
measurement or by the quality of the sample. This makes a part of the expected 
entropy change to become continuous and it shows up as a finite measurable 
specific heat. Interestingly, it is only very recently the expected features 
in specific heat in a first order transition has been seen experimentally in 
rare-earth compounds ~\cite{Gschneidner}. In the following discussion, we have 
to keep in mind the above special features since we are measuring specific 
heat near a transition which is expected to be first order. 
 
In figure 6,  $C_{exc}/T$ is plotted as a function of $T$ within a small 
interval of $\pm ~25K$ around $T_{CO}$ for both H = 0 and 8T. In figure
7, we have shown the excess entropy $S_{exc}(T)$ associated with this
transition. The entropy, $S_{exc}(T)$, for $T > 200 K$ is calculated by 
numerically integrating $C_{exc}/T$ as $S_{exc}(T) = 
\int_{200}^T C_{exc}/T dT$. The lower limit of the integral is purely 
a matter of convenience. It is far removed from $T_N$ and $T_{CO}$ and 
at T = 200K, $C_{exc} \approx$ 3$\%$ of $C_p$ and is negligible. 

We estimate the total change in entropy $\Delta S_{T}$ by using linear 
extrapolation of the entropy values from above and below to $T_{CO}$ as 
depicted in the fig. 7. $\Delta S_{T} \approx 1.8 J/mol~K$ at 0T and it 
decreases to $1.5 J/mol~K$ at 8T. This total change in entropy takes place 
over an interval of $\approx$ 10 -15K around $T_{CO}$. The magnitude of 
$\Delta S_{T} (0T)$ agrees well with that found in polycrystalline samples by
 Lees {\it et. al.} ~\cite{LEES} by integrating the area under the peak in 
$C_{exc}/T$. In polycrystalline $La_{0.35}Ca_{0.65}MnO_3$ Ramirez 
{\it et. al} ~\cite{Ramirez} obtained $\Delta S_T$$\approx 5 J/mole K$. 
In polycrystalline $Y_{0.5}Ca_{0.5}MnO_3$ our group had observed 
$\Delta S_T$ $\approx 2.5 J/mole K$ ~\cite{Arulraj}. The total entropy 
change at the CO transition thus seems to 
be a fraction of what one would expect from an order-disorder transition. 
The error in the estimation of the entropy change arising from the 
uncertainty in estimation of background $C_{Lattice}$ is not more 
than $5\%$. In a narrow temperature range over which the transition occurs 
the change in entropy being the difference of two quantities with similar 
background, the error arising from the estimation of the background is not 
severe.

The charge-ordered phase is expected to have a lower entropy than the 
charge-disordered insulating phase at $T > T_{CO}$. This implies that 
the sample absorbs latent heat to transform from the charge-ordered phase 
to the disordered phase. We would expect that atleast a part of the entropy
change $\Delta S_T$ is released as latent heat is absorbed from the 
total heat $\Delta Q$ supplied during the heat-pulse 
experiment and the sample temperature remains quite constant throughout 
the process. Since the change in sample temperature ($\Delta T$) is 
quite small, the specific heat $C_p = (dQ / dT)_p$ shows a very sharp 
peak near $T_{CO}$. As discussed before at a first-order transition 
$\Delta T \rightarrow 0$ and ideally $C_p$ should be a $\delta$ function 
at the transition temperature.
                                     
The peak however is broadened due to two reasons : \\
(a) Crystal quality : In most cases the expected $\delta$ function gets 
broadened by sample quality. The real crystal  contains some defects / 
inhomogeneities which would lead to broadening of the peak. With improved 
quality of the sample, the peak would be larger and sharper. For oxides 
containing multiple chemical constituents it is quite likely that such is 
the case. As shown in  figure 1, where we have compared the $C_p$ of a 
ceramic pellet and a single crystal, the single crystal has a much higher 
and narrower $C_p$ at $T_{CO}$ as compared to the ceramic sample.\\
(b) Measurement broadening : The second reason for broadening of the peak 
is connected to the measurement procedure itself. $C_p$ measurement 
involves measuring a finite temperature jump $\Delta T$ following the 
heat pulse $\Delta Q$. Since $C_p$ is well behaved in the region away 
from $T_{CO}$, it doesnot dependent much on the size of the temperature 
rise $\Delta T$ (as long as $C_p$ is not a very steep function of $T$). 
However, close to $T_{CO}$, the height as well as the width $\delta T$ 
of $C_p$ sensitively depends on $\Delta T$. \\

We have measured the latent heat by measuring $C_p$ with different 
$\Delta T$ as suggested by Gschneidner {\it et. al.} ~\cite{Gschneidner}. 
A sample of the data is shown in figure~8. We find that while away from
$T_{CO}$ there is no dependence of $C_p$ on the rise of $\Delta T$, as 
expected, the peak becomes narrower and higher when size of $\Delta T$ 
decreases as we approach $T_{CO}$. This continues till $\Delta T \le 0.5K$. 
For this value we reach a limiting width (at half maxima) $\delta T 
\approx 2K$. We believe that for $\Delta T \le 0.5K$, the width of the peak
is not determined by the measurement but by the crystal quality. This 
observation sets a limit to the height and width of $C_p$. We find from 
our measurement that the latent heat $L_{CO} / T_{CO} = \Delta S_{CO} 
\sim$ 1 J/mole K. Similar value of $\Delta S_{CO}$ was obtained for 
H = 8T. Our estimate of latent is a lower bound of the true latent heat 
and with a better crystal preferably with a transition width $< 1K$,
$\Delta S_{CO}/ \Delta S_T$ will increase substantially and may even 
$\rightarrow$ 1, as expected for a strong first order transition. 
We note that such a narrow peak might arise from a very narrow second 
order transition that has been broadened by defects or inhomogeneities. 
However we find that the applied field shifts the position of the peak but 
doesnot appreciably broaden the peak. This we take as a proof of a first 
order transition at $T_{CO}$ ~\cite{comment1}.

\subsection{Entropy change near $T_{MH}$ in 8T}
In a field of 8T the low temperature phase is an FMM or a spin-aligned metal, 
allowing for spin canting. But the phase is metallic as can be seen from 
figure~1. On heating, the FMM state (8T) becomes unstable towards the COI 
state formation and makes transition to the mixed charge ordered (MCO) state 
at $T_{MH}$. This is accompanied by a jump in $\rho$. We would like to ask if 
there is a change in entropy associated with the melting of the COI state to 
FMM state at $T_{MH}$ ? This question is interesting because both the phases 
have ordering of a kind, in addition the FMM phase is expected to have extra 
entropy due to presence of free electrons. A close look 
at the region $T \approx$ 95K, as shown in figure 9, shows a small 
dip in $C_p$ which as explained below can be considered as signature of this 
melting. It is important to note that, as expected in zero field, this feature 
is absent. As shown in fig. ~9, $C_p$(8T) starts to show change at $T = 
T_{MH} \approx$ 88K where $\rho$ shows a jump on heating and the specific 
heat transition is complete at $100K$ after a small dip in $C_{p}$, where 
the resistivity transition also stops. 

The small dip in $C_{p}$ is associated with a small heat release of 
$\approx$ 10J/mole at around 95K - 100K on heating. This suggests that on 
heating the FMM phase, stable in 8T for low T, starts to disorder. This 
probably destabilizes the FMM phase with regions of high resistivity COI 
regions appearing in it. Eventually at $T\approx T_{MH}$ the insulating 
regions increase in size and $\rho$ shows a jump in $\rho$. The resistivity 
transition has an element of percolation associated with it and thus occurs 
at a different volume fraction of the new phase in contrast to the specific 
heat transition which occurs mainly when the bulk of the sample is transformed.
Nevertheless, the small but distinct change in $C_p$ close to $T_{MH}$ is 
clearly seen.

In the temperature range $T_{MH} < T < T_{CO}$in a magnetic field(i.e, 
the MCO region) we find an interesting effect. The thermal 
relaxation time of the sample increases by more than one order of 
magnitude. This anomalously large relaxation is not observed at any 
temperature range in zero field or for $T>T_{CO}$ and $T<T_{MH}$ in 
$H=8T$. The existence of this large thermal relaxation necessitated 
that the sample be properly equilibrated before the data are taken. We 
have done that and the data presented here are taken after the sample 
has been properly equilibrated thermally. The thermal relaxation, 
however, is interesting in its own right and has been discussed in a 
separate publication ~\cite{Ayan3}.

\section{Low temperature $C_p$ in 0 and 8T magnetic field}
As it has been pointed out before, that the issue of low temperature specific 
heat ($T < 10K$) is controversial in manganites showing charge ordering, 
particularly in Pr$_{1-x}$Ca$_x$MnO$_3$ system. A large number of observations 
on low temperature specific heat is available for different compositions of 
Pr$_{1-x}$ Ca$_{x}$MnO$_{3}$, mostly in polycrystalline pellets and some in 
single crystals.

In figure~10 we show the  specific heat data of our sample with that of 
different compositions as obtained by different groups ~\cite{LEES,SMOLN}. 
Such a comparison is meaningful because it brings out the essential 
similarity and differences in these materials. A large linear term 
($\gamma \approx$ 30.6 mJ/mole-K$^2$ for x = 0.3 and $\gamma \approx$ 
15.7 mJ/mole-K$^2$ for x = 0.35 ~\cite{SMOLN}) 
appears in polycrystalline samples with x = 0.3 and x = 0.35 composition 
which decreases as x $\rightarrow$ 0.5. We note that such a large 
linear $\gamma$ has not been seen any of the single crystal data. From 
a comparison of the data  shown in figure~10 we can reach the following 
conclusions:\\
  (i) The polycrystalline ceramic samples  have higher specific heat than the 
single crystal samples, \\
 (ii) There is a clear trend that the specific heat of the ceramic samples 
decreases as we approach $x=0.5$.In particular,linear term in specific heat, is 
only observed for the ceramic samples and it decreases with increasing $x$.\\
(iii)The specific heat of the single crystal samples with compositions x = 0.37 
and 0.5 are very similar for $T~<~10K$.
 
In figure~11 we have compared the  specific heat of two close compositions: 
a polycrystalline x = 0.35 sample ~\cite{SMOLN} and a single crystal sample 
with x = 0.37 composition. The comparison is to elucidate how much of the 
excess specific heat is due to polycrystallinity and can we infer a likely 
origin of the excess specific heat. There are two mechanisms that can 
contribute to the excess specific heat: (1) Due to grains of small dimensions 
~\cite{Pohl} and (2)Due to two-level-systems(TLS) arising due to disorder 
as in amorphous solids and several disordered crystals ~\cite{Pohl}.
If the excess specific heat, is due to grains then $\delta C$ can be 
expressed as $\delta C = C_1 T + C_2 T^2$, where the coefficients $C_1$ 
and $C_2$ are related to the average grain diameter $< R >$ as $< R > = 
k_B [6 \hbar v_{sound} C_1]^{-0.5}$, where $v_{sound}$ is the sound velocity
in the crystal ~\cite{Pohl}. We can fit the experimentally observed 
$C_{poly}$ to the equation $C_{poly}=C_{sc}+ \delta C$. Using $C_{sc}$ 
from our observed data and the above expression of $\delta C$ we 
obtain an estimate of the average grain size $<R>$ and we obtain 
$<R> \approx  10^{-4} \mu m$  which is far too small compared to the 
typical $<R> >1 \mu m$ seen in most polycrystalline samples. We can thus 
rule out finite grain size as the source of excess specific heat.

If the excess specific heat $\delta C$ arises from TLS then it can be 
fit to a form $\delta C  = C_1 T + C_2 T^3$. This would arise 
from TLS with density of state $P(E) = \tilde a + \tilde b E^2$, 
where $\tilde a$ and $\tilde b$ are constants ~\cite{Pohl}. Using this 
expression for $\delta C$ we then fitted the 
observed $C_{poly}$ as had been done before. The fit is shown in figure 
~11. From the fitted parameters $C_1$ and $C_2$, we arrive at values of 
$\tilde a \approx$ 2.3 $\times$ 10$^{35}$ erg$^{-1}$/cc and     
$\tilde b \approx$ 1.5 $\times$ 10$^{35}$ erg$^{-3}$/cc. These values are 
quite comparable to but some what larger than those obtained in glasses. We 
conclude that extra specific heat of the polycrystalline samples  arise from 
excitations which behave as TLS. We have no clear understanding of the 
origin of these TLS but we can speculate that they will arise from 
incomplete orbital /charge order that can happen in polycrystalline 
samples due to random strains. What is needed for TLS to occur is to  
have multitude of ground states with almost degenerate energy. This   
can actually be due to the predominant incommensurate charge ordering 
as seen in both x = 0.35 and x = 0.37 compositions. As the charge-ordering 
becomes commensurate with the lattice with $x \rightarrow 0.5$, the 
difference in the specific heat between the polycrystalline and the 
single crystal samples decrease. We thus suggest that the disorder in 
polycrystalline samples give rise to TLS low energy excitations       
typically with energy $\leq 10K$ and as $x\rightarrow 0.5$, the $P(E) 
\rightarrow 0$.

Next we attend to the low temperature data taken on our single crystal 
sample. Figure~12 shows the specific heat data for the x = 0.37 sample
 plotted as $C_p/ T$ vs. $T^2$, a customary way to plot the data in 
anticipation of a linear term. It is clear that the linear term is absent 
in the COI state. This is not surprising for an insulating sample, the 
linear term was also absent in the specific heat data observed in the 
ceramic samples of $x = 0.4$ by Lees et. al ~\cite{LEES}. Neglecting the 
linear term, we have fitted the our observations to the following relation :
\begin{equation}
C_p = \alpha T^{-2} + \beta_3 T^3 + \beta_5 T^5
\end{equation}
where, $\alpha T^{-2}$ hyperfine contribution caused by the local magnetic 
field at the {\it Mn} nucleus due to electrons in unfilled shells, 
$\beta_3 T^3$ and $\beta_5 T^5$ are the lattice contribution to the 
specific heat, arising from phonons. A part of the $T^3$ contribution to 
$C_p$ can also arise from AFM spin waves.\\
The $T^3$ contribution is likely to be enhanced over and above the 
actual Debye contribution because of the presence of AFM spin waves, since 
our COI sample has canted AF order in this temperature range. The results 
of fitting the data to the above equation are shown in table 1. The 
parameters obtained by us are very similar to that seen by Lees {\it et. al}
~\cite{LEES}. In table 1 we have collected the parameters from different 
published data for comparison.

However, the interesting question is whether in a magnetic field of 8T 
where the COI is melted into a FMM phase, we see appearance of a linear 
specific heat term. In a past investigation on polycrystalline samples of 
La$_{0.5}$ Ca$_{0.5}$MnO$_3$ in a magnetic field of 8.5 T ~\cite{Smolyaninova} 
(which is high enough to melt the COI state) no linear term had been observed. 
The issue of absence of a linear term in the FMM phase obtained after melting 
the COI state is thus real. Particularly, when one compares it with the 
low temperature specific heat of the FMM phase as seen in the CMR region 
$\gamma~\approx$ 7.8 mJ/mole K$^2$ in La$_{0.8}$ Ca$_{0.2}$MnO$_3$ 
~\cite{Hamilton}, $\approx$ 5.2 mJ/mole K$^2$ in La$_{0.7}$ 
Ca$_{0.3}$MnO$_3$ ~\cite{Coey}. 

The $C_p / T$ vs. $T^2$ for our sample in presence of an 8T magnetic field
are shown in figure~12. The specific heat data have 
been fitted using the eqn. 1, with the parameters as shown in table 1. 
Interestingly, the specific heat in presence of magnetic field is distinctly 
lower than the specific in 0T for all $T < 40K$. Lowering of specific heat 
in presence of magnetic field has been reported Smolyaninova {\it et. al} 
~\cite{SMOLN}. Since resistivity measurements indicate the sample to be in 
a metallic state, one would expect an extra linear term appearing in the 
eqn.1 in presence of magnetic field. As seen in fig. 12, we couldnot detect 
the presence of any linear term even in presence of 8T. From the uncertainty 
in the data we find an upper limit of $\gamma \leq 0.1 mJ/mole K^2$, which is 
too small.

It is apparent from table 1, that contribution of the $\beta_3$ term is 
halved in presence of magnetic field. This appears to be the main  
reason for lower $C_p$ in $H = 8T$ compared to that in absence of magnetic 
field. This can be explained as follows. The  $T^3$ term contains an 
additional antiferromagnetic spin wave contribution for the COI sample. 
For an antiferromagnetic spin wave spectrum where $E = Dq$, the magnetic 
contribution to specific heat is $\propto T^3$. The melting of COI state 
to a FMM state leads to collapse of this AFM order and thus there will be 
no AFM spin wave contribution to the specific heat. This would result in 
a decrease of the $T^3$ term. However, presence of FM order should give 
a ferromagnetic spin wave contribution $\propto T^{3/2}$. We have 
attempted fitting a $T^{3/2}$ term to the 8T specific heat data.
However, we could not detect any $T^{3/2}$ term. Thus, it appears that 
the specific heat as observed in presence of magnetic field doesnot 
have a ferromagnetic spin wave contribution. 

Briefly, we find that the specific heat at low temperature needs more 
investigation in single crystalline CO systems and in high magnetic 
field. The existing data  (which includes too few single 
crystal data) both in zero field and magnetic field (high enough to 
melt the CO state) does not allow us to reach much definite conclusion 
other than the fact that in the FMM phase obtained by melting COI 
state in a high magnetic field does not have a large enough $\gamma$ 
that is comparable to FMM phases seen in manganites showing CMR 
behavior.

\section{Conclusion}
Our study of specific heat in a CO system done over an extended 
temperature range of 3K - 300K in presence of 0 and 8T magnetic 
field gives us a number of useful and new information regarding 
the thermodynamics of CO transition. This paper gives the first 
clear measurement of the entropy in a single crystal and also the 
latent heat (at T$_{CO}$) in zero field and in a magnetic field. The 
latent heat released proves the strong first order nature of the 
transition, which had never been shown to be so in actual measurement 
of specific heat. An estimate of the numbers like the latent heat 
in 0 and 8T field sets clearly the bound and scale within which 
any theoretical models must work. It is a very important number for 
any phase transition which gives the CO transition (both in 0T 
and 8T) a thermodynamic basis. Important is also the observation 
of a finite entropy change at the CO melting in a magnetic field. 
The result shows that a small but finite and qualitative difference 
in entropy exists between the FMM phase and the CI phase at the 
melting transition in the magnetic field where the former (FMM) 
phase (which is the low T phase) has a lower entropy than the 
other phase (which is the high T phase). Interesting also is the 
smallness of the entropy. 
The low temperature $C_p$ ($T < 15 K$), show no linear dependence 
on T as expected from an insulating sample. Surprisingly, this 
electronic term is absent even when the COI state is melted to 
a FMM state by a 8T magnetic field. An attempt had been made to 
understand the observed discrepancy between the specific heat 
of a single crystal and a polycrystalline sample in the low 
temperature regime. 

A.G. thanks the CSIR Center of Excellence in Chemistry, JNCASR, 
for financial support. A.K.R. thanks BRNS and DST for partial support 
through a sponsored scheme.

$^\dagger$ Electronic mail: arup@physics.iisc.ernet.in

\newpage
{\bf Figure captions :}
\vspace{0.5cm}

{\bf figure 1:} The temperature dependence of resistivity of Pr$_{0.63}$
Ca$_{0.37}$MnO$_3$ in presence of 0 and 8T magnetic fields.

{\bf figure 2:} $C_p$ of single crystalline Pr$_{1-x}$Ca$_{x}$MnO$_3$ 
(x = 0.37) over 3K- 300K in presence of 0 and 8T magnetic fields. Also 
plotted the $C_p$ of the polycrystalline x = 0.4 composition.

{\bf figure 3:} $C_p$ and $C_{exc}$ close to $T_N$ in H = 0T.

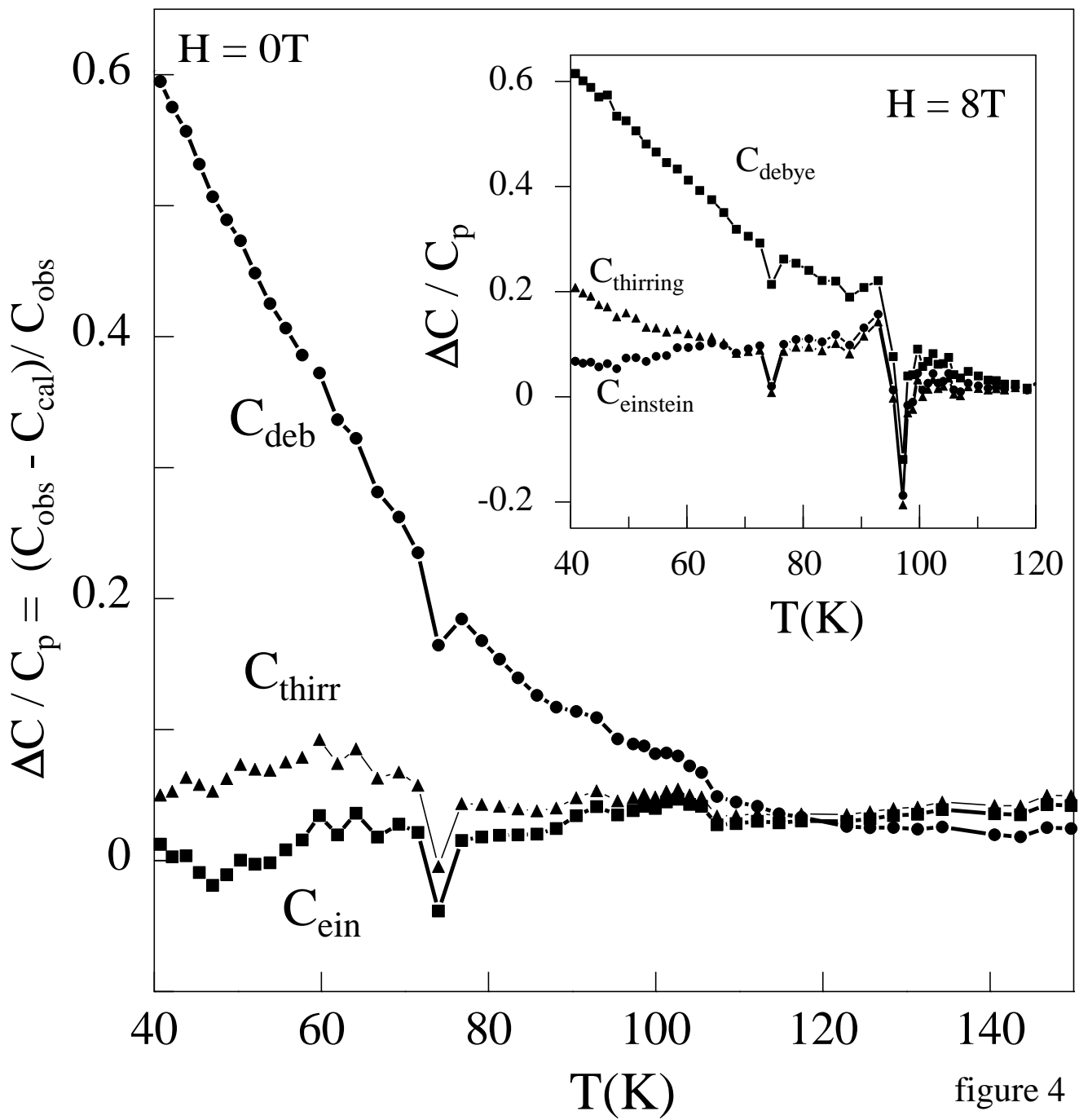
{\bf figure 4:} Deviation of $C_p$ from the fit for different models. 
Inset shows the same near $T_{MH}$ in H = 8T. 

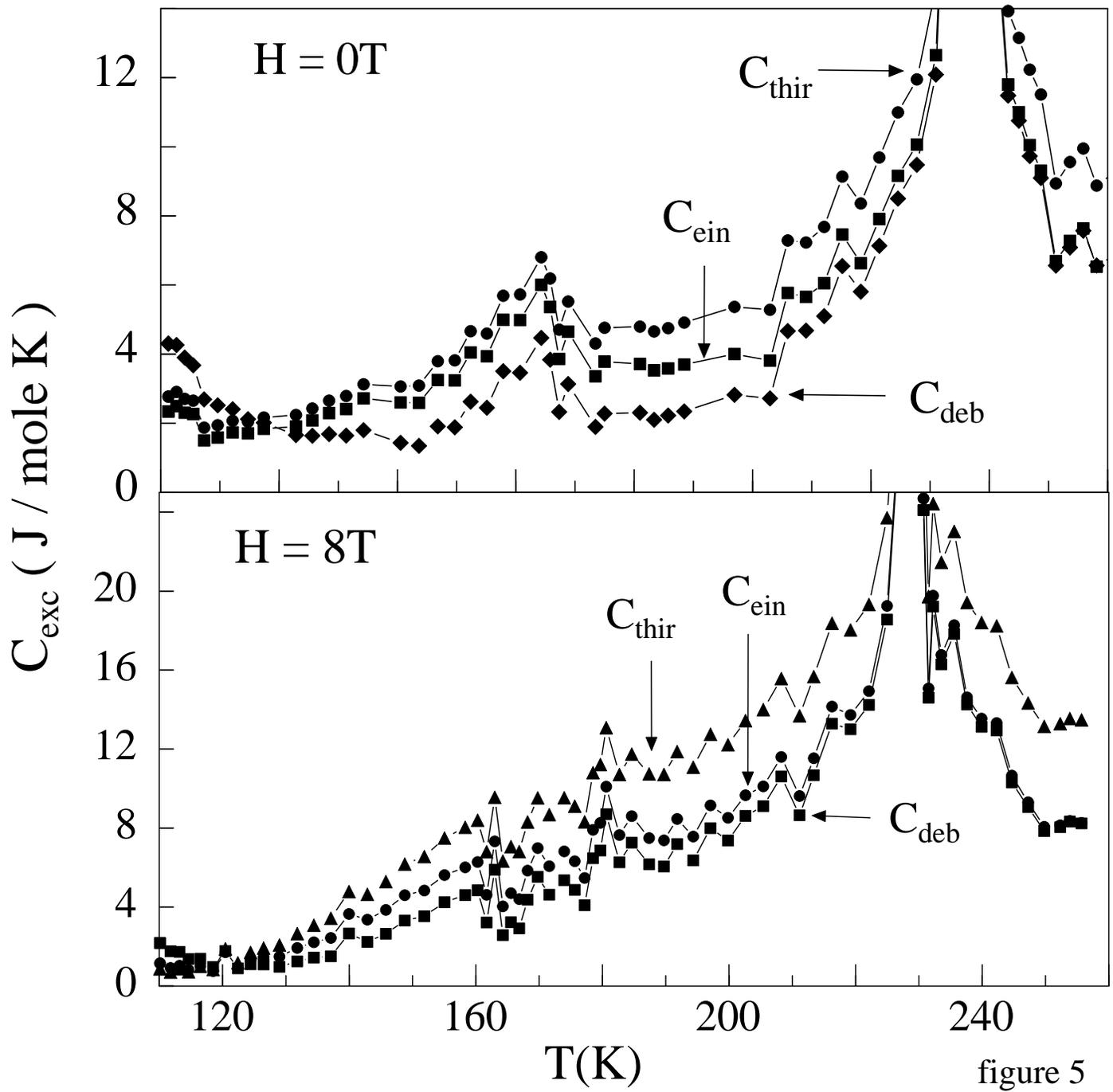
{\bf figure 5:} $C_{exc}$ obtained after subtracting the lattice 
background (as determined by the three models) for both H = 0 and 8T.

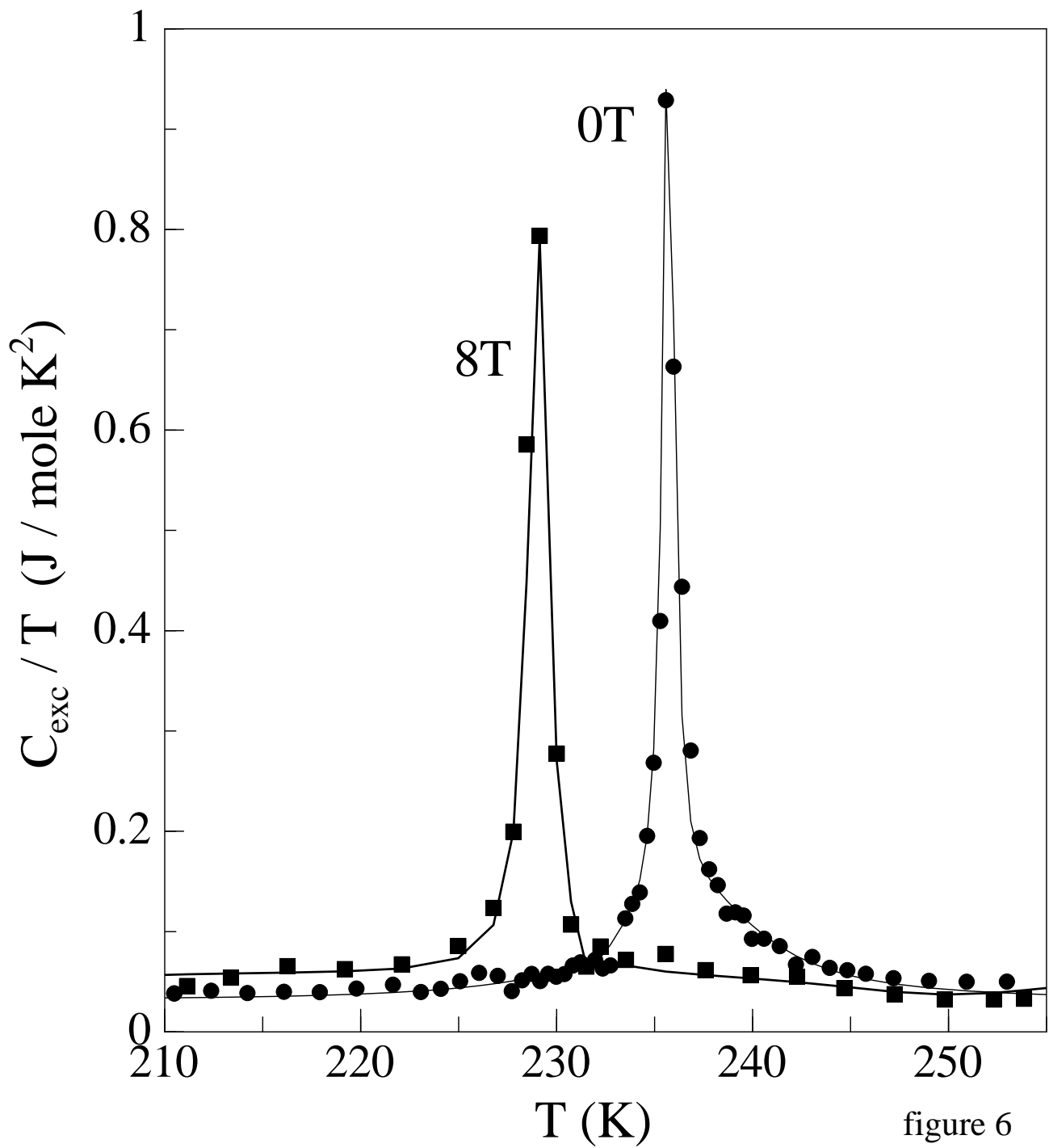
{\bf figure 6:} $C_{exc}/T$ around $T_{CO}$ in H = 0 and 8T.

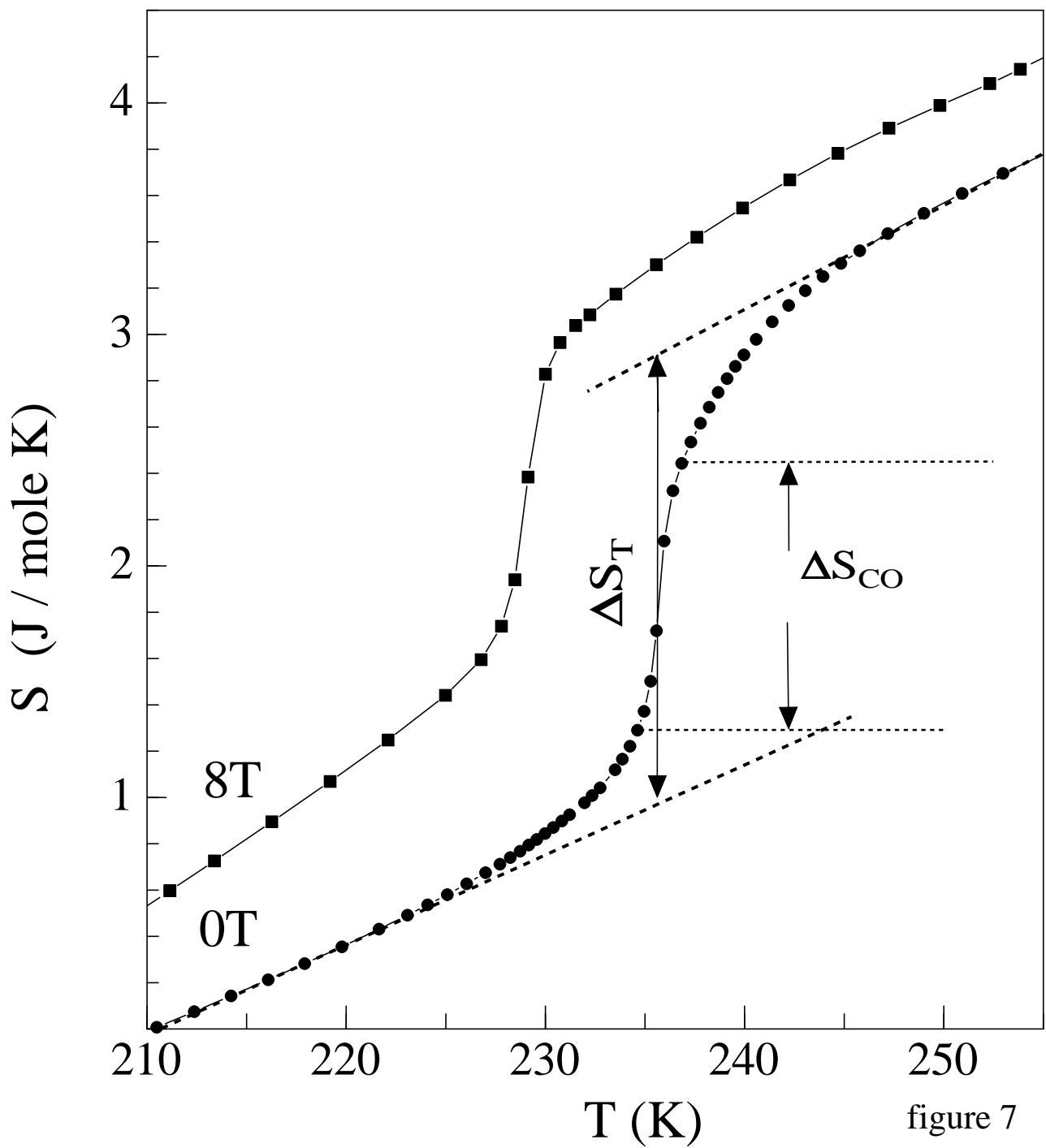
{\bf figure 7:} Excess entropy $S_{exc}$ as calculated from numerically
integrating $C_{exc}/T$ around $T_{CO}$ in H = 0 and 8T.

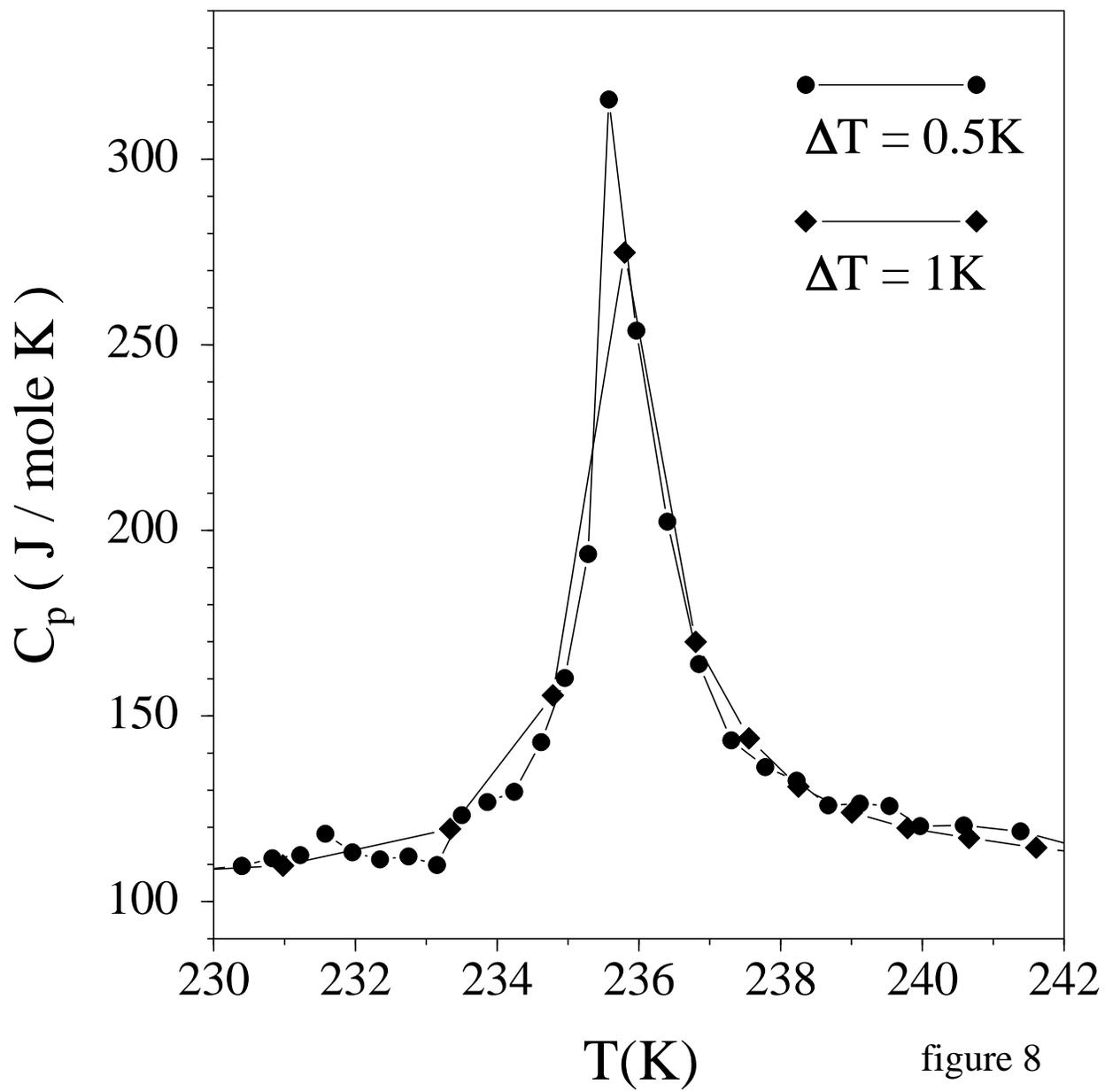
{\bf figure 8:} Varying sharpness and width of $C_p$ near $T_{CO}$ for
two different temperature rise $\Delta T$ = 0.5 and 1K.

{\bf figure 9:} $C_p$ and resistivity near the $T_{MH}$. The arrow marks
the melting of the COI state.

{\bf figure 10:} A comparison of $C_p$ of Pr$_{1-x}$Ca$_{x}$MnO$_3$ in 
the low temperature regime as observed by different groups, x = 0.5(sc),0.45(ce),0.35(ce)~\cite{SMOLN}, x = 0.4(ce) ~\cite{LEES} 
and x = 0.37(sc) is our sample.

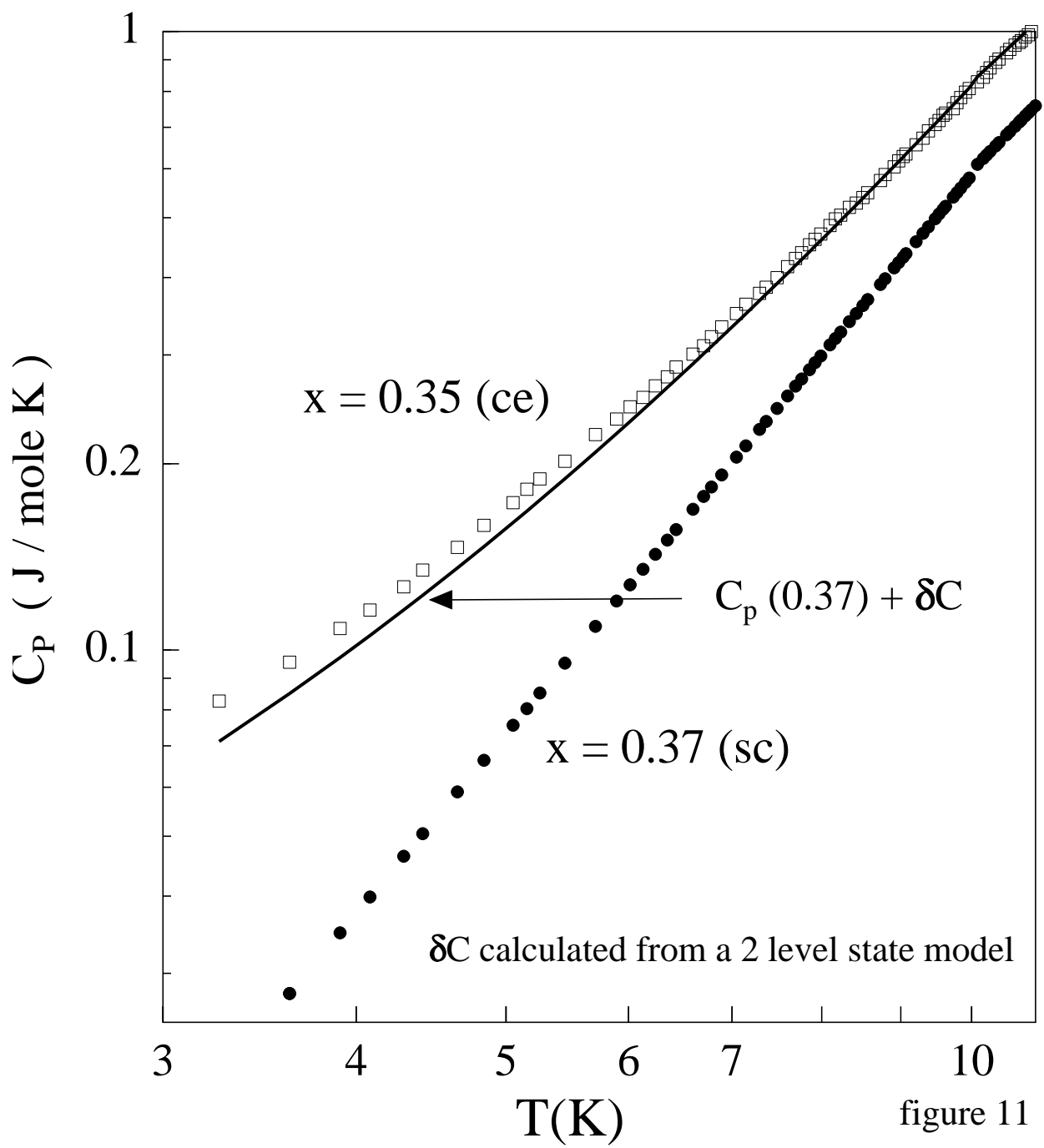
{\bf figure 11:} The excess specific heat, $\delta C$, as seen in the 
polycrystalline sample arising due to presence of two level states.  

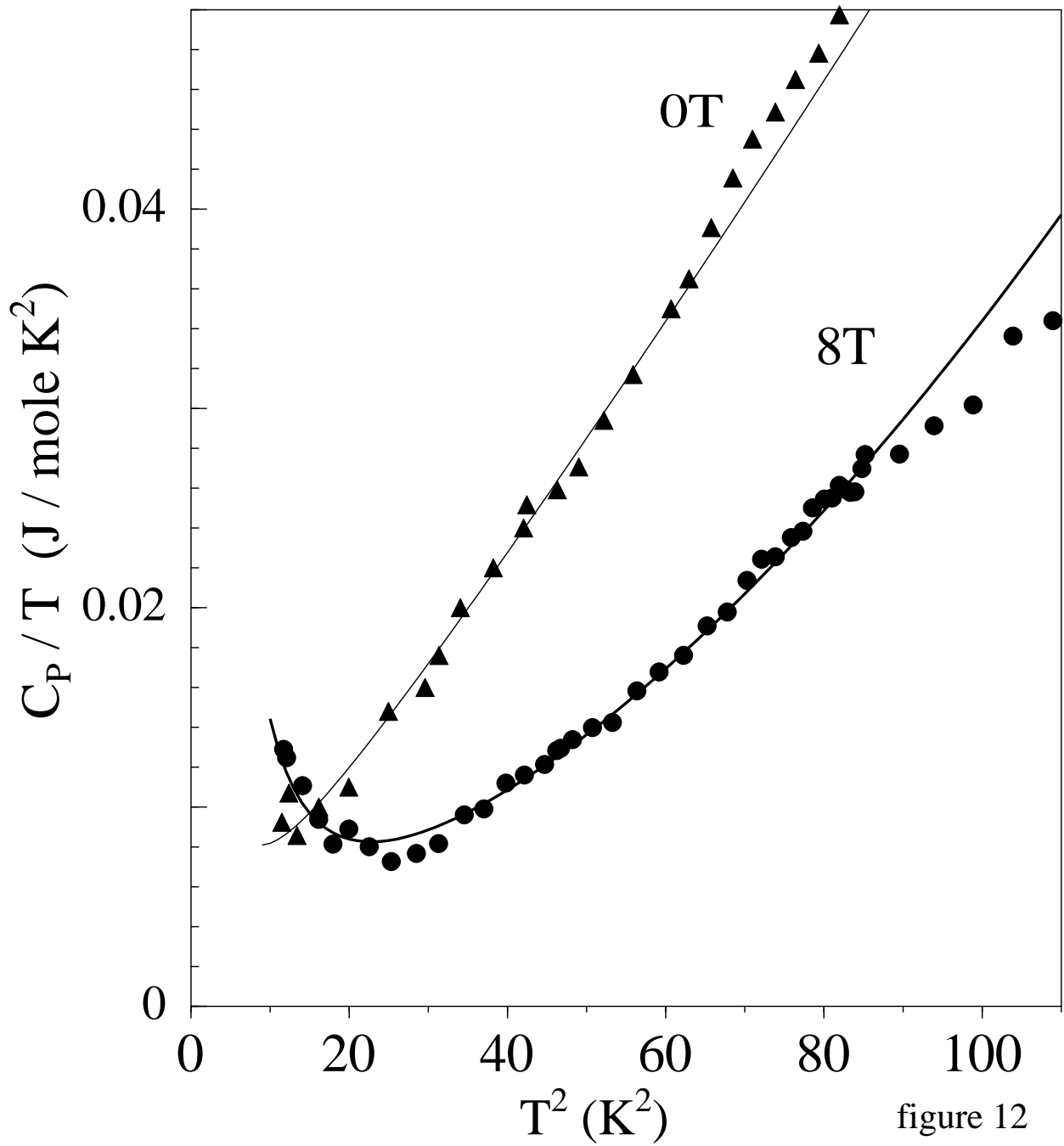
{\bf figure 12:} The low temperature $C_p$ of single crystalline
Pr$_{0.63}$Ca$_{0.37}$MnO$_3$ in H = 0 and 8T. The solid line is 
the fitted equation.

\begin{table}[tbp]
\caption{ The fitting results for the $C_p$ of Pr$_{1-x}$Ca$_x$
MnO$_3$ as obtained by various groups. The units of different quantities 
are : $\alpha$ (mJ K/mole), $\gamma$ (mJ/mole K$^2$), $\beta_3$ 
(mJ/mole K$^4$) and $\beta_5$ ($\times$ 10$^{-4}$ mJ/mole K$^6$)}

\begin{tabular}{ccccc}
x&$\alpha$&$\gamma$&$\beta_3$&$\beta_5$  \\
(Pr$_{1-x}$Ca$_x$MnO$_3$) &  &  &  &   \\ \hline
0.3\tablenotemark[1] & 63.0 & 30.6 & 0.30 &   \\
0.35\tablenotemark[1] & 56.0 & 15.7 & 0.39 &  \\
0.45\tablenotemark[1] & 28.0 & 3.1 & 0.31 &  \\
0.5\tablenotemark[1] & 22.0 & 2.4 & 0.26 &  \\
0.4\tablenotemark[2] & 28.0 &   & 0.54 & 4.5 \\
0.37 (H = 0T)\tablenotemark[3] & 85.0 &  & 0.54 & 4.5 \\
0.37 (H = 8T) & 400.0 &  & 0.16 & 18.0
\end{tabular}
\tablenotetext[1]{from Ref.~\cite{SMOLN}}
\tablenotetext[2]{from Ref.~\cite{LEES}}
\tablenotetext[3]{our sample}
\end{table}

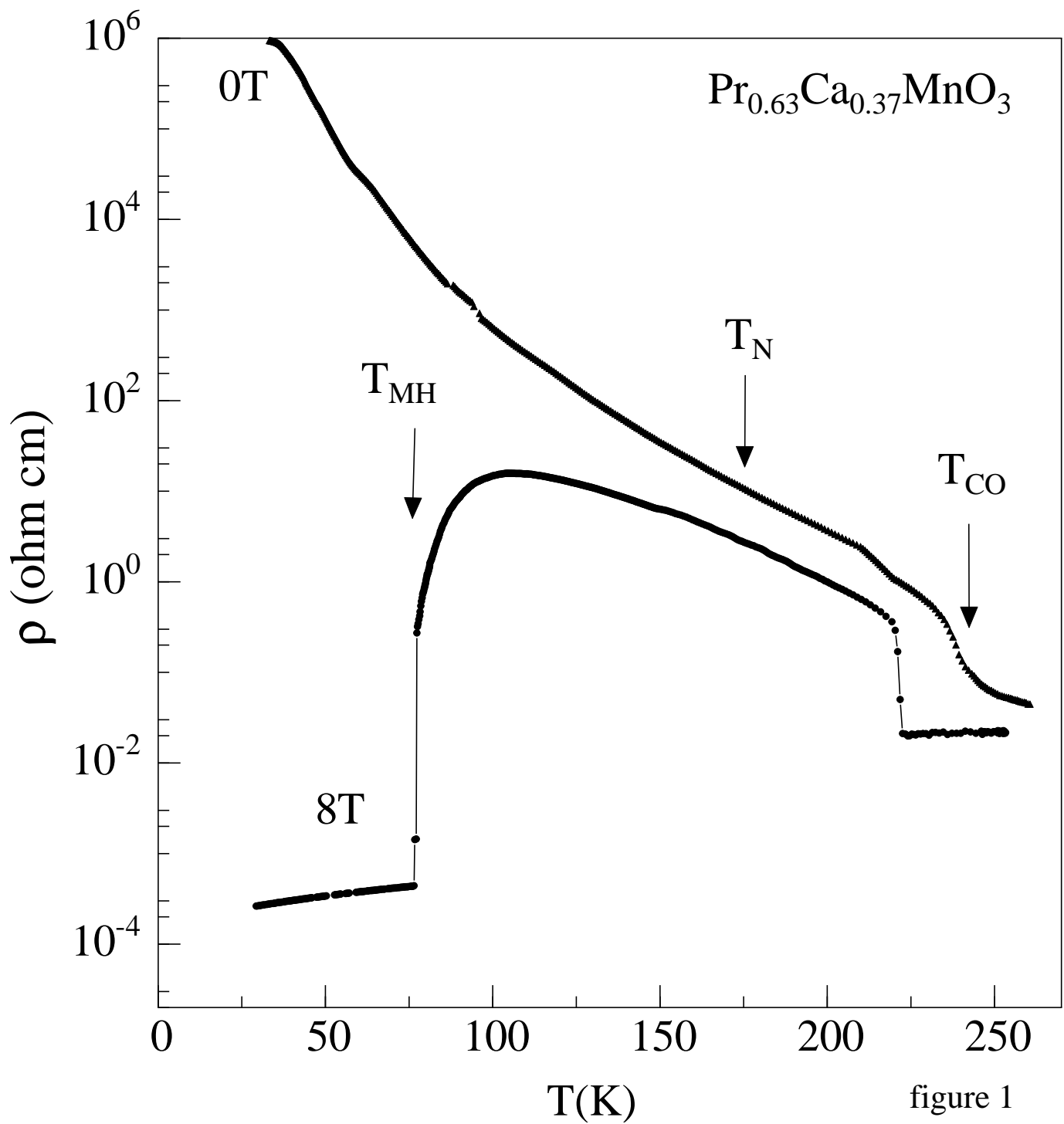
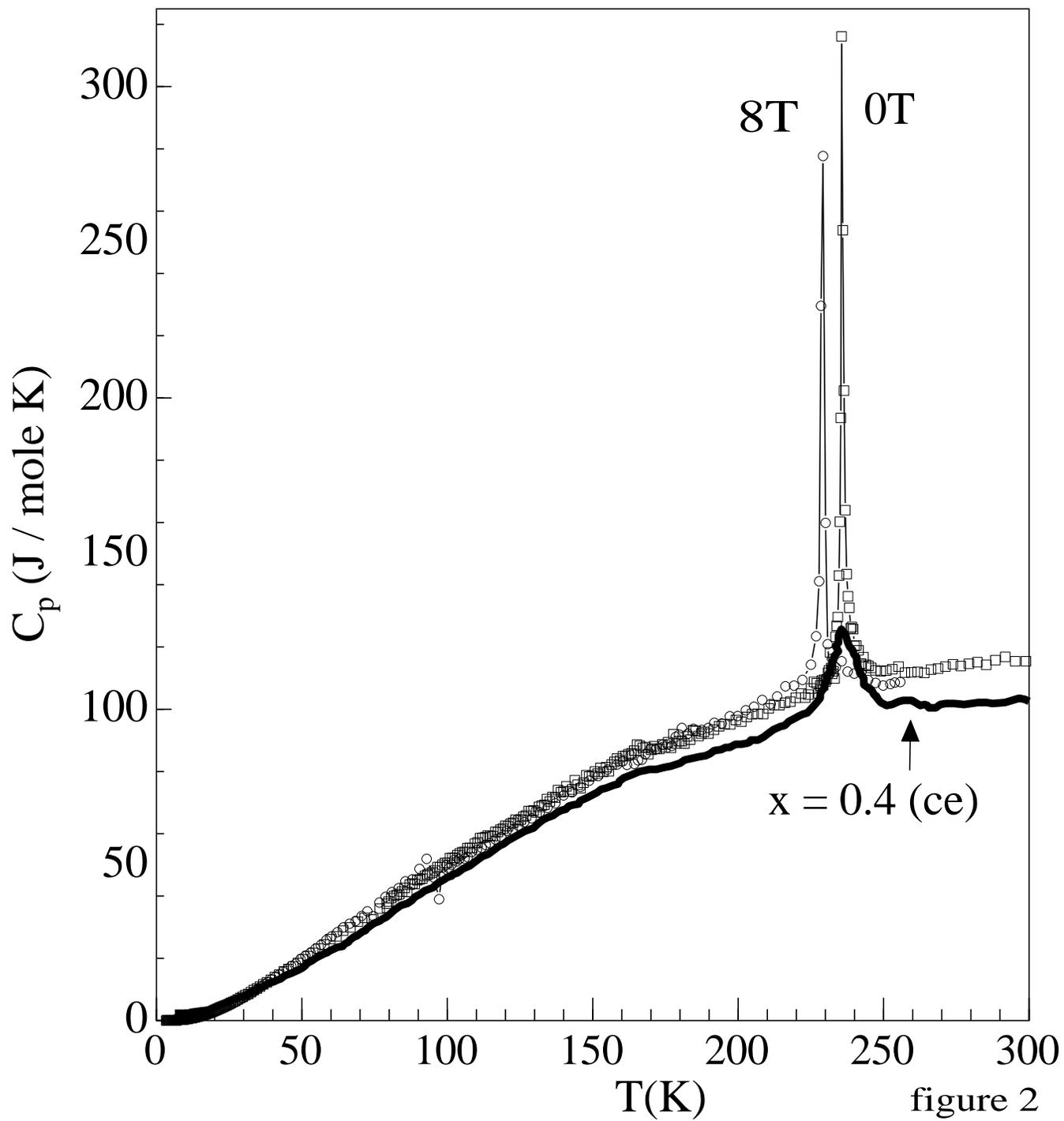
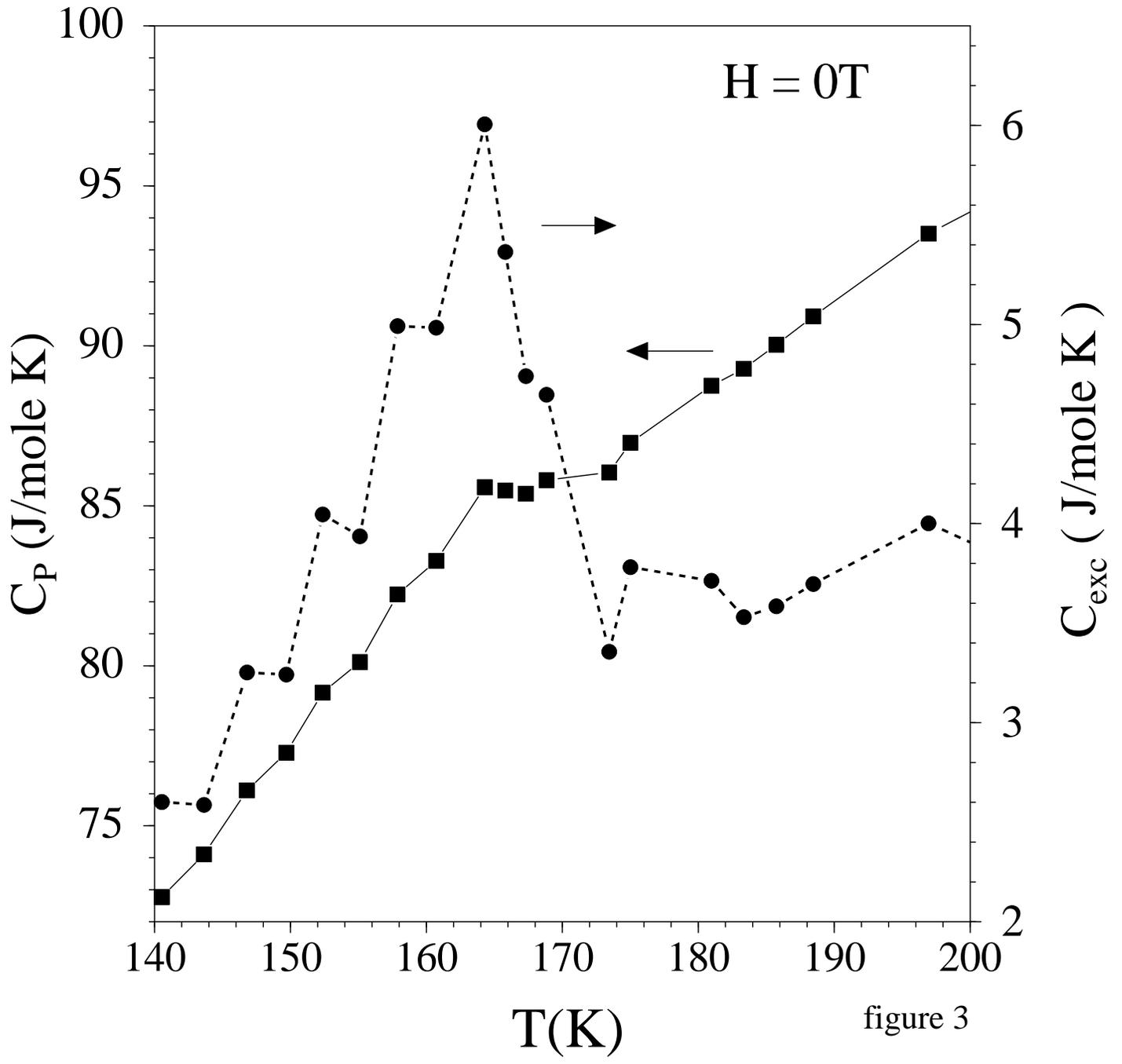
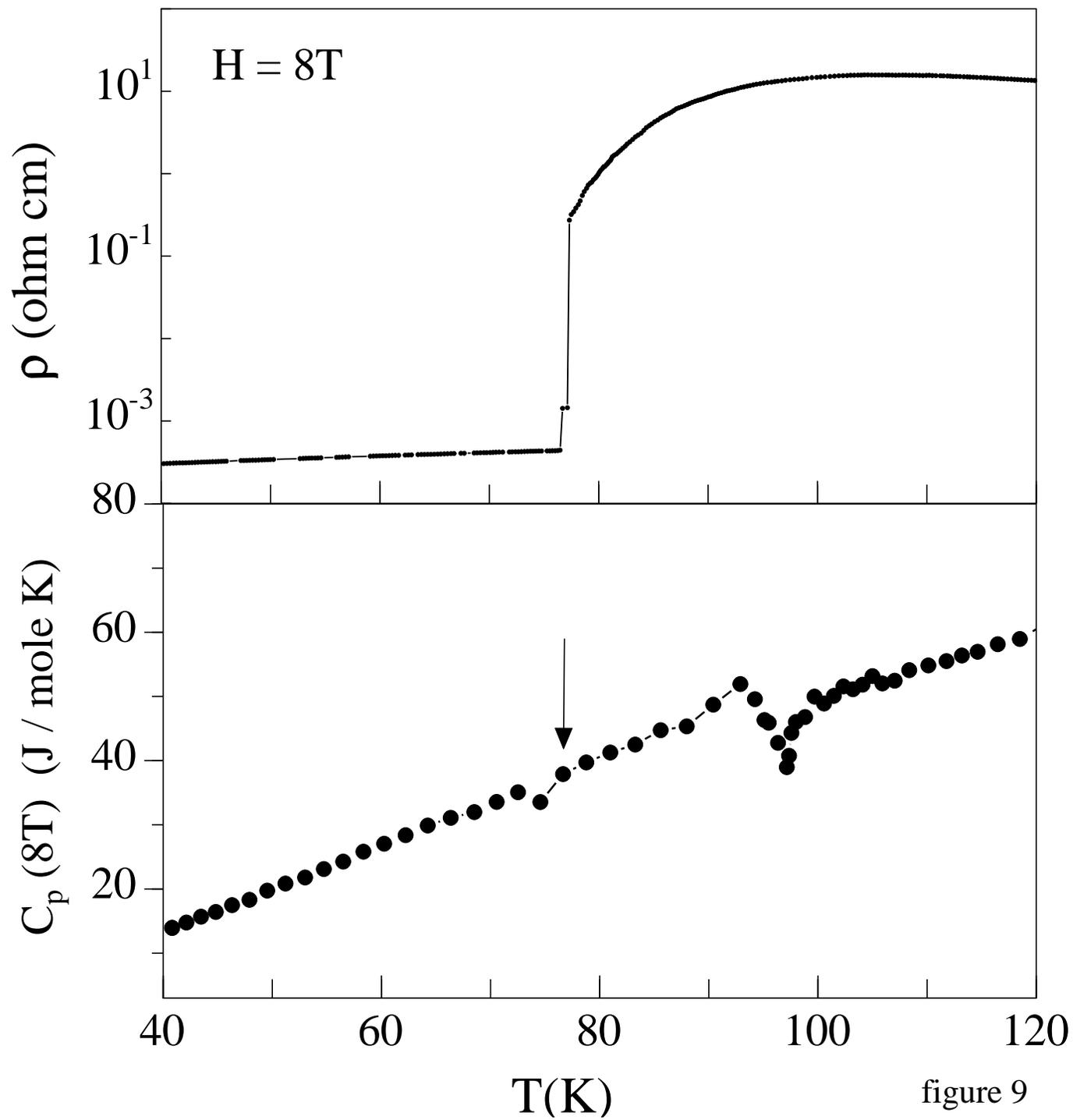
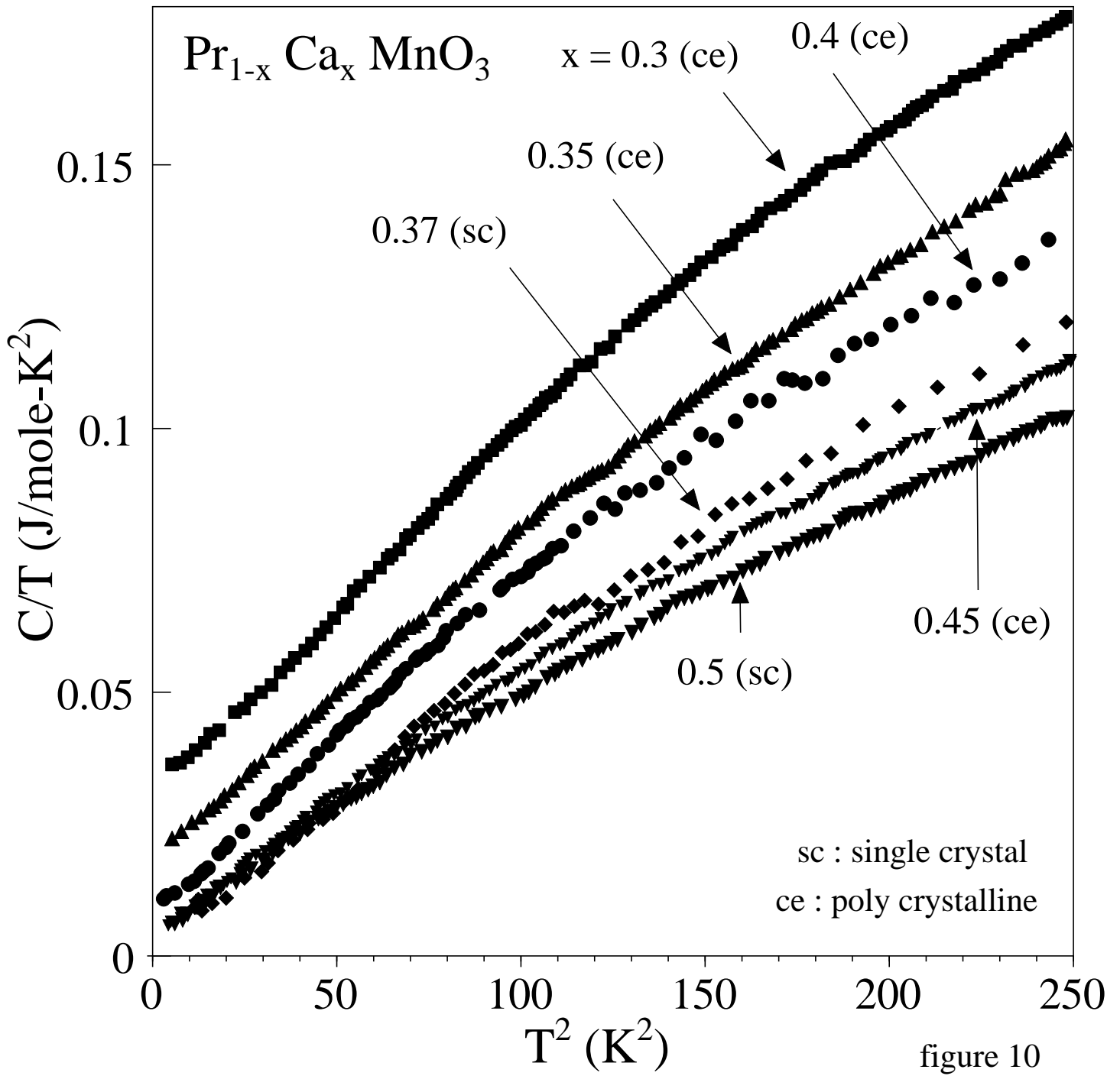
\end{document}